\documentclass[9pt]{article}
\usepackage{amsfonts}
\usepackage{pstricks}
\usepackage{pst-node}
\usepackage{epsfig}
\usepackage{epsfig}
\usepackage[latin5]{inputenc}
\usepackage[T1]{fontenc}
\usepackage{lipsum}
\usepackage{amsmath}
\usepackage{authblk}

\begin{document}
                \def\ba{\begin{eqnarray}}
                \def\ea{\end{eqnarray}}
                \def\w{\wedge}
                \def\d{\mbox{d}}
                \def\D{\mbox{D}}

%%%%%%%%%%%%%%%%%%%%%%%%%%%%%%%%%%%% TITLEPAGE %%%%%%%%%%%%%%%%%%%%%%%%%%%%%%
\begin{titlepage}
\title{Scattering of complex gravitino test fields from a gravitational pulse}
\author{Yorgo \c{S}eniko\u{g}lu\footnote{yorgosenikoglu@maltepe.edu.tr}}
\date{%
     \small Department of Basic Sciences, Faculty of Engineering and Natural Sciences, \\Maltepe University, 34857 Maltepe,\.{I}stanbul, Turkey\\%
}
%\affiliation{Department of Physics, Ko\c{c} University, 34450 Sar{\i}yer-\.{I}stanbul, Turkey }
\maketitle
%\vskip 1cm

%\date{4 September 2019}

%vskip 2cm

\begin{abstract}
Gravitational pulse waves are defined as step functions at the boundaries. In this setting, linearized Rarita-Schwinger equations are precisely solved in Rosen coordinates. It is found that the gravitino's energy-momentum varies with the sandwich wave's shape. The gravitino's energy density will decrease as it crosses the sandwich wave at the test field level since the background won't change.
 \end{abstract}

\vskip 1cm



\end{titlepage}

\newpage
%%%%%%%%%%%%%%%%%%%%%%%%%%%%%%%%%%%% INTRODUCTION %%%%%%%%%%%%%%%%%%%%%%%%%%%%%%

\section{Introduction}
An extremely brief explosion of a gravitational source may result in the formation of a gravitational sandwich wave, which travels through a region of the universe at the speed of light. Everything, even the weakly interacting neutrino, becomes susceptible to such gravitational waves since gravitational interaction is a universal phenomenon. In a previous work\cite{dereli-halilsoy}, we have studied the scattering of a neutrino in a gravitational sandwich wave spacetime at the test field level. It is discovered that the energy distribution exhibits significant fluctuations even at this level while crossing the sandwich wave. It has still to be determined if these variances will cause neutrinos to oscillate between different kinds. In that study, we looked into massless test neutrino fields in the background of sandwich plane waves, which might be either pure gravitational, pure electromagnetic, or mixed sandwich waves of both. A test neutrino experiences energy oscillations when it passes through a sandwich gravitational wave, depending on the values of the parameters. Additionally, the neutrino field's phase would change. Both of these phenomena are intriguing in and of themselves, and it should be explored further to see if they permit modern detection.

A similar problem exists for a gravitino field within the context of $N=2$ Extended Supergravity theory. The first supergravity theory to be constructed is the $N=1$ simple supergravity theory \cite{freedman-ferrara-vanN},\cite{deser-zumino}, based on the massless $(2,\frac{3}{2})$ supermultiplet, consisting of a graviton and a gravitino. Soon after, $N>1$ extended supergravity theories were found; the first being the $N=2$ model \cite{ferrara-vanN} which involves a $U(1)$ gauge field which couples non-minimally to a pair of massless gravitinos. Immediately, renormalisability properties \cite{deser-kay-stelle} were demonstrated and exact plane wave solutions were presented in the literature with \cite{finkelstein-kim}, \cite{dereli-tucker},\cite{hull}.

 The massless neutrino and massless gravitino have resemblances that need to be exploited; we will be perturbing spin $\frac{3}{2}$ fields. The direct coupling of the Einstein-Hilbert gravitational field to the spin $\frac{3}{2}$ Rarita-Schwinger field may result in the propagation of the gravitino field acausally. This is due to the works of Velo and Zwanziger \cite{velo-zwanziger}; according to their findings, the Rarita-Schwinger field seems to propagate in these conditions acausally in the sense that the equation's properties are spacelike, enabling information about the field configuration to flow at speeds faster than the speed of light. Therefore to dispose of this inconsistency, we implement the supergravity theory in which the locally supersymmetric ansatz provides a foundation to a coherent theory. The field equations obtained are similar, but it turns out that you couple it to the Einstein-Maxwell theory. If we study the gravitational pulse in the simple supergravity case the model does not work, it gives a constraint to the metric functions, therefore we need to study this problem in the $N=2$ extended supergravity.

The paper is organised as follows. Section 2 is an overview of N=2 extended supergravity theory, while Section 3 presents the sandwich wave spacetime. Section 4 is the ansatz and solution of the scattering of a complex gravitino field crossing the gravitational pulse. The paper is concluded in Section 5.

\bigskip

%%%%%%%%%%%%%%%%%%%%%%%%%%%%%%%  SECTION 2 %%%%%%%%%%%%%%%%%%%%%%%%%%%%%%%%%%%%%%
%\newpage

\section{N=2 Extended Supergravity Theory}

In the $N=2$ extended supergravity theory, a pair of spin $\frac{3}{2}$ gravitino fields are coupled non-minimally and locally supersymmetrically to the Einstein-Maxwell theory. Using the variational principle, the field equations for the theory may be obtained from the action \cite{beler-dereli1},
\ba
I[e,\omega,\psi^k, F] = \int_M {\mathcal{L}};
\ea
the Lagrangian density 4-form explicitly is \cite{beler-dereli2}
\ba
{\mathcal{L}} &=& \frac{1}{2} R_{a b} \wedge *(e^a \wedge e^b) + \frac{i}{2} \bar{\psi}^k \w \gamma_5\gamma \w D\psi^k - \frac{1}{2}F \w *F +\frac{i\sqrt{2}}{4}\epsilon^{jk}(\bar{\psi}^j \w \psi^k)\w *F \nonumber \\
  &-&\frac{i\sqrt{2}}{4}\epsilon^{jk}(\bar{\psi}^j \w \gamma_5\psi^k)\w F + \frac{1}{16}\epsilon^{ij}(\bar{\psi}^i \w \psi^j) \w \epsilon^{kl}*(\bar{\psi}^k \w \psi^l)\nonumber \\
  &-&\frac{1}{16}\epsilon^{ij}(\bar{\psi}^i \w \psi^j) \w \epsilon^{kl}(\bar{\psi}^k \w \gamma_5\psi^l).
\ea
The orthonormal basis 1-forms used to define the spacetime metric are represented by the $e^a$'s
\ba
g = \eta_{ab} e^a \otimes e^b = -e^0 \otimes e^0 + e^1 \otimes e^1 + e^2 \otimes e^2 + e^3 \otimes e^3.
\ea
* is the Hodge dual with the spacetime orientation $*1=e^0\w e^1\w e^2\w e^3$. $\omega_{ab}$ are the metric compatible orthonormal connection 1-forms, $\gamma=\gamma_a e^a$ and $\psi^k$'s are 2 real spinor valued 1-forms
\ba
\psi^k=\psi^k \hspace{0.01mm}_a \hspace{0.2mm}e^a, \quad k=1,2.
\ea
A global $SO(2)$ symmetry defined as $\epsilon^{12}=-\epsilon^{21}=1, \epsilon^{11}=\epsilon^{22}=0$ is referenced by the latin indices $i$, $j$, and $k$.

\noindent
We define the external covariant derivative as
\ba
D\psi=d\psi + \frac{1}{2}\omega^{ab}\sigma_{ab} \w \psi,
\ea
where
\ba
\sigma_{ab}=\frac{1}{4}[\gamma_a,\gamma_b].
\ea
Finally, $F$ is defined as
\ba
F=\frac{1}{2}F_{ab}e^a \w e^b.
\ea
Since the couplings are non-minimal, one does not have to introduce a potential 1-form, but using Lagrange multipliers we can impose the constraint that $dF=0$. Even if $dF=0$ implies locally that $F=dA$, it does not have to be global so we will use $F$ as our main variable.
The action density's variational field equations are as follows:
\ba
-\frac{1}{2}R^{bc} \w *(e_a \w e_b \w e_c) &=& -\frac{i}{2} \bar{\psi}^k \w \gamma_5\gamma_aD\psi^k-(F_{ac}F^{c}\hspace{0.01mm}_b+\frac{1}{4}\eta_{ab}F_{cd}F^{cd})*e^b \nonumber \\
&+&\frac{i\sqrt{2}}{2}[\epsilon^{kj}(\bar{\psi}^k_a\psi^j_c)F^c\hspace{0.01mm}_b+\epsilon^{kj}(\bar{\psi}^k_b\psi^j_c)F^c\hspace{0.01mm}_a+\frac{1}{2}\eta_{ab}\epsilon^{kj}(\bar{\psi}^k_c\psi^j_d)F^{cd}]*e^b\nonumber \\
&+&\frac{1}{2}[\epsilon^{kj}(\bar{\psi}^k_a\psi^j_c)\epsilon^{il}(\bar{\psi}^i_c\psi^l_b)+\frac{1}{4}\eta_{ab}\epsilon^{kj}(\bar{\psi}^k_c\psi^j_d)\epsilon^{il}(\bar{\psi^c}^i{\psi^d}^l)]*e^b,
\ea
which are the Einstein equations;
\ba
i\gamma_5\gamma \w D\psi^k-\frac{i}{2}T^a \w \gamma_5\gamma_a\psi^k&=&\frac{i\sqrt{2}}{2}\epsilon^{kj}\gamma_5\psi^j \w F -\frac{i\sqrt{2}}{2}\epsilon^{kj}\psi^j \w *F \nonumber \\
&-&\frac{1}{4}\epsilon^{kj}\psi^j \w \epsilon^{il}*(\bar{\psi}^i \w \psi^l)+\frac{1}{8}\epsilon^{kj}\gamma_5\psi^j\w \epsilon^{il}(\bar{\psi}^i \w \psi^l)\nonumber \\
&+&\frac{i}{8}\epsilon^{kj}\psi^j\w \epsilon^{il}(\bar{\psi}^i \w \gamma_5\psi^l),
\ea
which are the Rarita-Schwinger equations.

The remaining equations are Maxwell's equations and the algebraic expression that determines the torsion 2-forms are
\ba
d*F=\frac{i\sqrt{2}}{4}\epsilon^{kj}d*(\bar{\psi}^k \w \psi^j)-\frac{i\sqrt{2}}{4}\epsilon^{kj}d(\bar{\psi}^k \w \gamma_5\psi^j)
\ea
and
\ba
T^a=\frac{i}{4}\bar{\psi}^k \w \gamma^a\psi^k.
\ea

\medskip
Our study is at the linearized level and we are not considering any back reaction, so the field equations at the test field level become:
\ba
-\frac{1}{2}\hat{R}^{bc} \w *(e_a \w e_b \w e_c) &=&-(F_{ac}F^{c}\hspace{0.01mm}_b+\frac{1}{4}\eta_{ab}F_{cd}F^{cd})*e^b,\nonumber \\
i\gamma_5\gamma \w \hat{D}\psi^k&=&\frac{i\sqrt{2}}{2}\epsilon^{kj}\gamma_5\psi^j \w F -\frac{i\sqrt{2}}{2}\epsilon^{kj}\psi^j \w *F,\nonumber \\
d*F&=&0,
\ea
where the hat denotes the use of the Levi-Civita connections.
We remark at this point that bosonic parts are equivalent to Einstein-Maxwell equations; nevertheless the gravitino equations are non-trivial and give a mixing $SO(2)$ matrix on the right hand side. Consequently we can take any Einstein-Maxwell solution and solve the linearized Rarita-Schwinger equations.
%%%%%%%%%%%%%%%%%%%%%%% SECTION 3  %%%%%%%%%%%%%%%%%%%%%%%%%%%%%
%\newpage
\section{Sandwich Wave Spacetime}
The Brinkmann form of the gravitational or electromagnetic pp-wave metric is
\begin{equation}\label{Brinkmann}
g=2dUdV + 2 H(U,X,Y)dU^2 + dX^2 + dY^2.
\end{equation}

A straightforward extension of shock or step waves, sandwich waves may be described in terms of Heaviside step functions that have a non-zero value across the limited range $0 \leq U \leq U_0$. In this situation, the wave-front area ahead of $U = 0$ and the region behind it $U=U_0$ are both described by the flat Minkowski metric. In the Einstein-Maxwell theory, a generic form of a sandwich plane wave metric  can be found in \cite{halilsoy}. One may easily create pure gravitational, pure electromagnetic, or combinations of the two valid for the finite time curvature zone by taking the proper constraints. The gravitational $\psi_4$ and electromagnetic $\phi_{22}$ components of a linearly polarized plane sandwich wave are thus found to be independent of the coordinates $X$ and $Y$ that span the transverse plane. Then $H(U,X,Y)$ is expressed as
\begin{equation}
  H(U,X,Y)=\frac{1}{2}[(\Theta(U)-\Theta(U-U_0)][a^2(X^2+Y^2)-b^2(X^2-Y^2)],
\end{equation}
where the gravitational and electromagnetic parameters, respectively, are $a$ and $b$. The following coordinate transformation is typically believed to be more practical for using Rosen's metric form \cite{rosen} to illustrate the transverse characteristic of such spacetimes.
\ba
U=u, \quad U_0=u_0, \\
V=v - \frac{1}{2}(x^2ff_u+y^2hh_u), \\
X=xf, \quad Y=yh
\ea
so that the metric becomes
\begin{equation}\label{Rosen}
  g=2dudv + f(u)^2 dx^2 + h(u)^2 dy^2.
\end{equation}
The solutions to the field equations are given as
\ba
f(u)=cos[A(u\theta(u)-(u-u_0)\theta(u-u_0))]-Asin(Au_0)(u-u_0)\theta(u-u_0), \nonumber \\
h(u)=cos[B(u\theta(u)-(u-u_0)\theta(u-u_0))]-Bsin(Bu_0)(u-u_0)\theta(u-u_0).
\ea
where $A^2=(a^2-b^2)$ and $B^2=(a^2+b^2)$.
The gravitational wave background will be examined in this work for one instance that initially corresponds to a pure electromagnetic sandwich wave. By setting the proper restrictions, it is possible to deduce the pertinent quantities in this situation from the generic answer that was presented previously.
Figure 1 shows the overall plan for thinking about such a gravitational sandwich wave geometry. The metric functions $f(u)=h(u)=1$ and the related metric are provided by in the flat Minkowski area $u\leq0$.
\begin{equation}
  g=2dudv + dx^2 + dy^2.
\end{equation}
When $0\leq u \leq u_0$, the second region is a non-trivial curved region with a sandwich wave given by the metric
\begin{equation}
  g=2dudv + f(u)^2dx^2 + h(u)^2dy^2.
\end{equation}
Recall that the metric functions $f(u)$ and $h(u)$ are related to the sandwich wave's particular background geometry. Finally, Region III, which is once again flat, is characterized by $u>u_0$. However, unlike the first region, this is not stated explicitly. In reality, the third region's metric is provided by
\begin{equation}\label{region3}
  g=2dudv + \bar{f}(u)^2dx^2 + \bar{h}(u)^2dy^2,
\end{equation}
The following modification can be used to reveal the metric's (\ref{region3}) flat nature
\ba
U&=&u, \nonumber \\
X&=&x\bar{f}, \quad
Y=y\bar{h}, \nonumber \\
V&=&v-\frac{1}{2}x^{2}\bar{f}\bar{f}_u-\frac{1}{2}\bar{h}\bar{h}_u y^{2}
\ea
that produces
\begin{equation}
ds^{2}=2dUdV+dX^{2}+dY^{2}.
\end{equation}

\begin{figure}
  \centering
  \includegraphics[width=1.00 \textwidth]{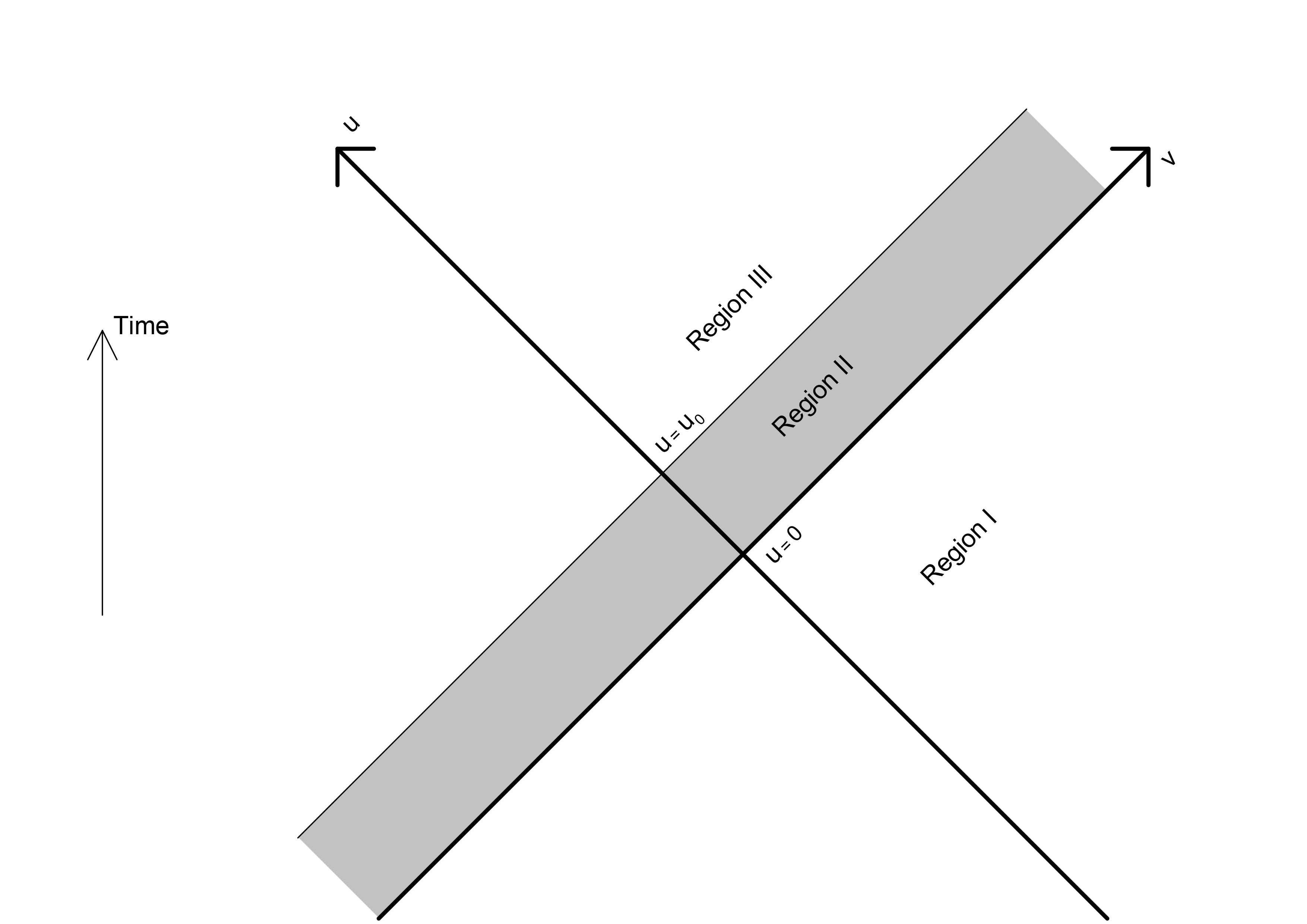}
  \caption{The gravitational sandwich wave geometry's overview. The front of the sandwich wave is represented by Region I, a flat Minkowski region. The Sandwich wave is found in Region II, which is curved. The sandwich wave's back is Region III, again a flat Minkoski region.}
\end{figure}

It is necessary to patch together these three independent regions-two flat ones outside and one curved within the limited duration plane fronted wave region-by using the proper junction conditions. Bell and Szekeres \cite{bell-szekeres} have demonstrated that the O'Brien and Synge junction conditions, which demand the continuity of the metric $g_{\mu\nu}$ and $g^{ij}g_{ij,u}$, are appropriate junction conditions for colliding plane electromagnetic waves in Einstein-Maxwell theory. This enables us to avoid any source term across the boundaries. The global sandwich wave solution in Einstein-Maxwell theory has been shown \cite{halilsoy} to satisfy the O'Brien and Synge junction criteria across the two borders $u=0$ and $u=u_0$.

\section{Rarita-Schwinger equation in a Sandwich Wave background}
The curvature zone is represented by the non-trivial metric
\begin{equation}\label{Rosen2}
  g=2dudv + f(u)^2 dx^2 + h(u)^2 dy^2.
\end{equation}
Let's start by introducing the complex null basis 1-forms for convenience as
\ba
l=\frac{1}{\sqrt{2}} (e^3+e^0), \quad n=\frac{1}{\sqrt{2}} (e^3-e^0), \quad m=\frac{1}{\sqrt{2}}(e^1+ie^2),
\ea
then we have
\ba
l=du,\quad n=dv, \quad m=\frac{1}{\sqrt{2}}(fdx+ihdy)
\ea
and the metric is
\ba
g=2l \otimes n + 2m \otimes \bar{m}.
\ea
$\gamma$ can also be expressed as
\ba
\gamma=\gamma_ul + \gamma_vn+\gamma_{-}m+\gamma_{+}\bar{m}
\ea
where
\ba
\gamma_u=\frac{1}{\sqrt{2}} (\gamma_3+\gamma_0), \quad \gamma_v=\frac{1}{\sqrt{2}} (\gamma_3-\gamma_0), \quad \gamma_{+}=\frac{1}{\sqrt{2}} (\gamma_1+i\gamma_2), \quad \gamma_{-}=\frac{1}{\sqrt{2}} (\gamma_1-i\gamma_2).
\ea
Let's now build the gravitino ansatz. We will create a form that simply takes into account the physical modes and depicts progressive waves moving in the $\frac{\partial}{\partial v}$  direction. Therefore our ansatz will have the form
\ba
\psi^k=\eta^kl+\xi^k\bar{m}+\xi^{*k} m
\ea
where $\eta^k(u,\zeta,\bar{\zeta})$ and $\xi^k(u,\zeta,\bar{\zeta})$ are two real and two complex spinors, $\zeta$ being the stereographic projection coordinate, with $\gamma_v\eta^k=\gamma_v\xi^k=0$.

By assuming that the gravitino field be transverse, which is assured by
\ba
d*\psi^k=0
\ea
and traceless by
\ba
*\gamma \w \psi^k =0,
\ea
our ansatz will equivalently verify the conditions $\xi^k,_\zeta=0$ and $\gamma_{-}\xi^k=0$.

\medskip
The coupled Rarita-Schwinger equations to be solved in this background characterized by $f(u)=h(u)=cos(au)$ and $F=a l\w m+ a l \w \bar{m}$ are
\ba
\frac{i\gamma_5}{\sqrt{2}f}(\gamma_{-}\eta^1,_{\bar{\zeta}}-\gamma_{+}\eta^1,_{\zeta})&=&\frac{ia\gamma_5}{\sqrt{2}}(\xi^2-\xi^{*2})-\frac{a}{\sqrt{2}}(\xi^2+\xi^{*2}) \nonumber \\
\frac{i\gamma_5}{\sqrt{2}f}(\gamma_{-}\eta^2,_{\bar{\zeta}}-\gamma_{+}\eta^2,_{\zeta})&=&-\frac{ia\gamma_5}{\sqrt{2}}(\xi^1-\xi^{*1})+\frac{a}{\sqrt{2}}(\xi^1+\xi^{*1})
\ea
We choose a real representation of the $\gamma$ matrices as
\ba
\gamma_0 =
\begin{pmatrix}
i\sigma_2 & 0 \\
0 & -i\sigma_2
\end{pmatrix}, \quad
\gamma_1=
\begin{pmatrix}
\sigma_1 & 0 \\
0 & \sigma_1
\end{pmatrix}, \quad
\gamma_2=
\begin{pmatrix}
0 & -i\sigma_2 \\
i\sigma_2 & 0
\end{pmatrix}, \quad
\gamma_3=
\begin{pmatrix}
\sigma_3 & 0 \\
0 & \sigma_3
\end{pmatrix}.
\ea
Let us write explicitly $\eta^k$ and $\xi^k$
\ba
\eta^1=
\begin{pmatrix}
Re\delta_{1} \\
Re\delta_{2} \\
Re\delta_{3} \\
Re\delta_{4}
\end{pmatrix}, \quad
\eta^2=
\begin{pmatrix}
Re\beta_{1} \\
Re\beta_{2} \\
Re\beta_{3} \\
Re\beta_{4}
\end{pmatrix}, \quad
\xi^1=
\begin{pmatrix}
\alpha_1^{*} \\
\alpha_2^{*} \\
\alpha_3^{*}\\
\alpha_4^{*}
\end{pmatrix}, \quad
\xi^2=
\begin{pmatrix}
\epsilon_1^{*} \\
\epsilon_2^{*} \\
\epsilon_3^{*} \\
\epsilon_4^{*}
\end{pmatrix},
\ea
where $\alpha_i, \beta_i, \delta_i$ and $\epsilon_i$ are odd complex valued functions, $Re$ and $Im$ stand for real and imaginary part respectively.

\medskip
\noindent
The transverse and traceless conditions for these functions imply that
\ba
\xi^1=
\begin{pmatrix}
\alpha^{*} \\
\alpha^{*} \\
-i\alpha^{*}\\
i\alpha^{*}
\end{pmatrix}, \quad
\xi^2=
\begin{pmatrix}
\epsilon^{*} \\
\epsilon^{*} \\
-i\epsilon^{*} \\
i\epsilon^{*}
\end{pmatrix},
\ea
and the solutions to the coupled Rarita-Schwinger equations are found to be
\ba
\eta^1=
\begin{pmatrix}
Re\delta \\
Re\delta \\
-Im\delta \\
Im\delta
\end{pmatrix}, \quad
\eta^2=
\begin{pmatrix}
Re\beta \\
Re\beta \\
-Im\beta \\
Im\beta
\end{pmatrix},
\ea
with $\beta=\beta(u), \hspace{2mm} \delta=\delta(u), \hspace{2mm} \alpha=\alpha(u,\bar{\zeta})$ and $\epsilon=\epsilon(u,\bar{\zeta})$.

\medskip
Analyzing the gravitino solutions shown above will help one comprehend the nature of the effect of a sandwich gravitational wave spacetime on a test gravitino field. It is essential for that goal to determine the elements of the gravitino stress-energy-momentum tensor. The gravitino stress-energy-momentum tensor's orthonormal parts $T_{ab}$ may be used to define the gravitino stress-energy 3-forms $\tau_a = T_{ab}*e^b$.
By plugging the functions above in the energy momentum expression, the explicit form of the energy density is found to be:
\ba
T_{00}=\frac{1}{2}\big(\alpha^*\alpha_u-\alpha_u^*\alpha+\epsilon^*\epsilon_u-\epsilon_u^*\epsilon + 2(\left|\epsilon\right|^2+\left|\alpha\right|^2)\frac{f_u}{f}\big).
\ea
Assuming that $\alpha$ and $\epsilon$ do not exhibit a fundamental change during the process between $u=0$ and $u=u_0$, the shift in the energy density of the gravitino $\Delta\rho$ is defined as
\ba
\Delta\rho=\rho_{out}-\rho_{in}=(\left|\epsilon\right|^2+\left|\alpha\right|^2)\frac{f_u}{f}.
\ea
Explicitly for our case
\ba
\Delta\rho=-a(\left|\epsilon\right|^2+\left|\alpha\right|^2)tan(au_0).
\ea
We note that since $a>0$ and $u_0$ is small, the gravitino loses energy when it is scattered through the gravitational pulse.
%%%%%%%%%%%%%%%%%%%%%%%%%%%%%   CONCLUSION    %%%%%%%%%%%%%%%%%%%%%%%%%%%%%%%%%%

%\newpage

\bigskip

\section{Conclusion}
A test gravitino field in a gravitational sandwich wave background space-times was explored. In Rosen coordinates, a particular type of gravitational wave called a sandwich wave has a curvature that is non-zero only across the limited area $0\leq u\leq u_0$. Both the front and the back of such sandwich waves favor the Minkowski metric. In these three regions, one can find precise background solutions to the Rarita-Scwhinger equation.  An important remark to be made at this point is that in the $N=2$ extended supergravity, $F$ is not the electromagnetic $F$. Although the 2-form $F$ that we explore here is a $U(1)$ gauge field and the action is the Maxwell action, it is not in fact the electromagnetic $F$, as it couples to the complex supergravity field non-minimally. In the ordinary Einstein-Maxwell cases a minimal coupling for the Faraday tensor is expected, but there exist a non -minimal coupling here. Therefore $F$ should be the massless, spin-1, "graviphoton field" that induces repulsive gravitational interactions. $(\psi^1,\psi^2)$ is an $SO(2)$ doublet and $SO(2)$ is isomorphic to $U(1)$. Since the couplings are non-minimal, a possible potential 1-form was not necessary. Looking at the test field level, the bosonic part of the theory is found to be identical to the Einstein-Maxwell theory, therefore a solution was found using the Bell-Szekeres solution of the Einstein-Maxwell theory. We remark that the metric function enters in a non-trivial way in the gravitino's energy momentum-tensor and therefore in the energy density. We deduce from energy considerations that the gravitino's energy shifts in the sense that it has less energy after crossing the gravitational sandwich wave.

\bigskip

\section{Acknowledgement}
I am indebted to Prof. Tekin Dereli for the encouragement and insightful discussions regarding this study.

\bigskip
%\newpage
\section*{Data Availability Statement}
No new data were created or analysed in this study.


\begin{thebibliography}{99}
\bibitem {dereli-halilsoy} {T.Dereli, M.Halilsoy, O.Gurtug, Y.Senikoglu}{\sl Neutrino fields in a sandwich gravitational
wave background},Class.Quant.Grav.{\bf 39}(2022) 225018
\bibitem{freedman-ferrara-vanN} D.Freedman,S.Ferrara,P.van Nieuwenhuizen,{\sl Progress toward a theory of supergravity}, Phys.Rev.{\bf D13}(1976)3214
\bibitem{deser-zumino} S.Deser,B.Zumino,{\sl Consistent supergravity},Phys.Lett.{\bf 62B}(1976) 335
\bibitem{ferrara-vanN} S.Ferrara,P.van Nieuwenhuizen,{\sl Consistent Supergravity with Complex Spin-3/2
 Gauge Fields}, Phys.Rev. Lett.{\bf 37}(1976)1669
\bibitem{deser-kay-stelle} S.Deser,J.H.Kay,K.S.Stelle,{\sl Renormalisability properties of supergravity},Phys.Rev.Lett.{\bf 38}(1977)527
\bibitem{finkelstein-kim} R.J.Finkelstein,J.Kim,{\sl Classical solutions of the equations of supergravity},J.Math.Phys.{\bf 22}(1981)2228
\bibitem{dereli-tucker} T.Dereli,R.W.Tucker,{\sl A class of exact supergravity solutions},Phys.Lett.{\bf 97B}(1980)396
\bibitem{hull} C.M.Hull,{\sl Killing spinors and exact plane-wave solutions of extended supergravity},Phys.Rev.{\bf D30}(1984)334
\bibitem{velo-zwanziger} G.Velo, D.Zwanziger {\sl Propagation and quantization of Rarita-Schwinger waves in an external
electromagnetic potential}, Phys. Rev. {\bf 186}(1969)1337-41
\bibitem{brinkmann} H.Brinkmann, {\sl Einstein spaces which are conformally mapped on each other}, Math.Ann.{\bf 94} (1925) 119
\bibitem{halilsoy} M.Halilsoy,{\sl Test field in a sandwich wave spacetime}, Class. Quant. Grav. {\bf 14} (1997) 2231
\bibitem{rosen} N.Rosen, Phys.Z.Sov.U.{\sl On gravitational waves}, {\bf 12} (1937) 366
\bibitem{bell-szekeres} P. Bell and P.Szekeres,{\sl Interacting Electromagnetic Shock Waves in General Relativity}, Gen. Rel. Grav.{\bf 5} (1974) 275

%\bibitem{salam-sezgin} {\bf Supergravities in Diverse Dimensions: Commentary and Reprints}. Edited by A.Salam,E.Sezgin in two volumes.\\(North-Holland and World Scientific,1989)

%\bibitem{bern et al} Z.Bern,J.Carrasco,W.-M.Chen,A.Edison,H.Johansson,J.Parra-Martinez,\newline R.Roiban,M.Zheng,{\sl Ultraviolet properties of $\mathcal{N}=8$ supergravity at 5-loops}, Phys.Rev.{\bf D98}(2018) 086021
%\bibitem{dereli-aichelburg} P.C.Aichelburg,T.Dereli,{\sl Exact plane wave solutions of supergravity field equations},Phys.Rev.{\bf D18}(1978)1754
%\bibitem{dereli-aichelburg2} P.C.Aichelburg,T.Dereli,Phys.Lett.{\bf 80B}(1978)357
%\bibitem{urrutia} L.F.Urrutia,{\sl A new exact solution of classical supergravity},Phys.Lett.{\bf 102B}(1981)393
\bibitem{beler-dereli1} A.Beler,T.Dereli,{\sl Plane waves in supergravity},Class.Q.Grav.{\bf 2}(1985)147
\bibitem{beler-dereli2} A.Beler,T.Dereli,{\sl Plane waves in N=2 extended supergravity},Class.Q.Grav.{\bf 2}(1985)823

%\bibitem{volkov-galtsov} M.S.Volkov,D.Gal'tsov,{\sl Finite-dimensional Grassmann  algebras and structure of the equations of simple supergravity},Theo.Math.Phys.{\bf 80}(1989)716
%\bibitem{embacher1} F.Embacher,{\sl A new class of exact pp-wave solutions in simple supergravity},J.Math.Phys.{\bf 25}(1984)1484
%\bibitem{beler-dereli2} A.Beler,T.Dereli,Class.Q.Grav.{\bf 2}(1985)823
%\bibitem{rosenbaum} M.Rosenbaum,M.Ryan,L.F.Urrutia,R.A.Matzner,{\sl Colliding plane waves in $N=1$ classical supergravity},Phys.Rev.{\bf D34}(1986)409
%\bibitem{tucker} R.W.Tucker,{\sl Affine transformations and the geometry of superspace},J.Math.Phys.{\bf 22}(1981)422
%\bibitem{benn} I.M.Benn,{\sl Complex quaternionic formulation of $SL(2,C)$ gauge theories of gravitation}, unpublished Ph.D.thesis,University of Lancaster,1981
%\bibitem{benn-tucker} I.M.Benn,R.W.Tucker,{\sl Extended Lorentz invariance and field theory},J.Phys,{\bf A 14}(1981)1745
%\bibitem{morita} K.Morita,{\sl Quaternions and simple $D=4$ supergravity},Prog.Theo.Phys.{\bf 72}(1984)1056
%\bibitem{morita2} K.Morita,{\sl Quaternionic variational formalism for Poincar\'{e} gauge theory and supergravity},Prog.Theo.Phys.{\bf 73}(1985)999


\end{thebibliography}
\end{document}